# Window of Opportunity for Mitigation to Prevent Overflow of ICU capacity in Chicago by COVID-19

*Sergei Maslov and Nigel Goldenfeld*

*Carl R. Woese Institute for Genomic Biology, University of Illinois at Urbana-Champaign*

*March 18 2020*

**Please note: this is a working document and has not been submitted for journal publication. It is planned that a later version of this document will be submitted for peer-reviewed publication, but in the interests of sharing information during a rapidly changing epidemic landscape, we are making this early version available.**

**Executive Summary**

We estimate the growth in demand for ICU beds in Chicago during the emerging COVID-19 epidemic, using state-of-the-art computer simulations calibrated for the SARS-CoV-2 virus. The questions we address are these:

(1) Will the ICU capacity in Chicago be exceeded, and if so by how much?

(2) Can strong mitigation strategies, such as lockdown or shelter in place order, prevent the overflow of capacity?

(3) When should such strategies be implemented?

Our answers are as follows:

(1) The ICU capacity may be exceeded by a large amount, probably by a factor of ten.

(2) Strong mitigation can avert this emergency situation potentially, but even that will not work if implemented too late.

(3) If the strong mitigation precedes April 1st, then the growth of COVID-19 can be controlled and the ICU capacity could be adequate. The earlier the strong mitigation is implemented, the greater the probability that it will be successful. **After around April 1 2020, any strong mitigation will not avert the emergency situation.** In Italy, the lockdown occurred too late and the number of deaths is still doubling every 2.3 days. It is difficult to be sure about the precise dates for this window of opportunity, due to the inherent uncertainties in computer simulation. But there is high confidence in the main conclusion that it exists and will soon be closed.

Our conclusion is that, being fully cognizant of the societal trade-offs, there is **a rapidly closing window of opportunity to avert a worst-case scenario in Chicago**, but only with strong mitigation/lockdown implemented in the next week at the latest. **If this window is missed, the epidemic will get worse and then strong mitigation/lockdown will be required after all, but it will be too late.**

**Methods**

This document describes the results of computer simulations of a standard population level epidemiological model (SEIR-model) with seasonal affects and parameters calibrated to be appropriate for SARS-CoV-2. The calculations are done by solving differential equations of the SEIR model [1], without spatial extension, demographic stochasticity or attention paid to small-world and scale-free network effects, but these are potentially important and[2, 3] could be readily added in the future [4-8]. The model additionally has categories for severely sick people who are hospitalized, people in critical condition in need of ICU rooms and ventilators, and a fatal category. The simulation uses severity assumptions as a function of individual age, informed by epidemiological and clinical observations in China [9]; no modifications have been to take into account national differences, such as number of smokers in the population etc.

The model has been calibrated with the hospital data in the Chicago area and is able to account for the rapid rise in COVID-19 patients that we are starting to see (see Figure 1). Here is how we performed calibration. The simulation needs a starting assumed number of cases, and we initially used the value 100. Here is why. Previously we had estimated that the Chicago community cluster was 1600 infected individuals on March 14 2020. The way we did that estimate was to work backwards from the number of ICU confirmed covid-19 patients at a major Chicago hospital (5 confirmed, 110 persons under investigation (PUIs) but 10% of those will turn out to be covid-19 positive). Then we used the doubling time of 2.3 days which was true at that time, and used a 20% hospitalization rate for covid-19 patients, with a time interval between infection and hospitalization of 9.9 days. Then we found that this cluster originated around February 19 2020.

However, we were not satisfied with this argument, because it predicted the first covid-19 confirmed case would be too late. To do this, we experimented with moving the starting time of the simulation with one infected individual, and found that this would work if the date of infection was February 16$^{th}$ 2020. Reassuringly, this date is roughly consistent with the working backwards calculation! We ran the simulations with this initial condition: one infection event on Feb 16$^{th}$ 2020. Starting with this value, our simulations predicted that on March 18 2020, between 38 and 40 people would be hospitalized, with 5 patients in ICU. At that time, the data available indicated at least 18 ICU patients (of which 11 were a cluster at a single hospital) and at least 19 non-critical hospitalized patients. We regarded this agreement as satisfactory within the uncertainties inherent in computer simulation at the population level with such small numbers of cases.

Now we need to comment about the $R_0$ parameter used in the simulations. There is no real consensus on the right value for $R_0$ at this stage in the infection, so we experimented to find a value that replicated the observed doubling time of about 2.1-2.5 days in Illinois (Figure 1). We found that we could match the emergence of the infection in Illinois with an $R_0$ = 4.0 (annual average), latency = 5 days, infectious period = 3 days, seasonal forcing strength 0.2 and seasonal peak in January. This value is within the ranges reported in an extensive epidemiological study [10].

# Results

We ran the simulation for a town with a population of 2.71 million, and with age structure appropriate for the US. This is supposed to represent Chicago (not the Chicago Metropolitan Area). We calibrated the model to match the observed epidemiological statistics. We found that the "patient zero" who started the local cluster was infected around Feb 16 2020. The parameters of the model were selected to reproduce the observed doubling time of 2.0 days for reported cases in Illinois (Figure 1). The model predicts a current situation of 40 hospitalized non-critical patients, and 5 patients in ICU across Chicago. From a sample of three large hospitals in Chicago there are at least 19 hospitalized non-critical patients and additional 18 in ICU. Due to the inevitable uncertainties in computer modeling and the chance events that dictate early infections, we regard this agreement as satisfactory. The most important thing is that our simulations are calibrated to reproduce well the growth trend, so that we can predict the future course of the epidemic.

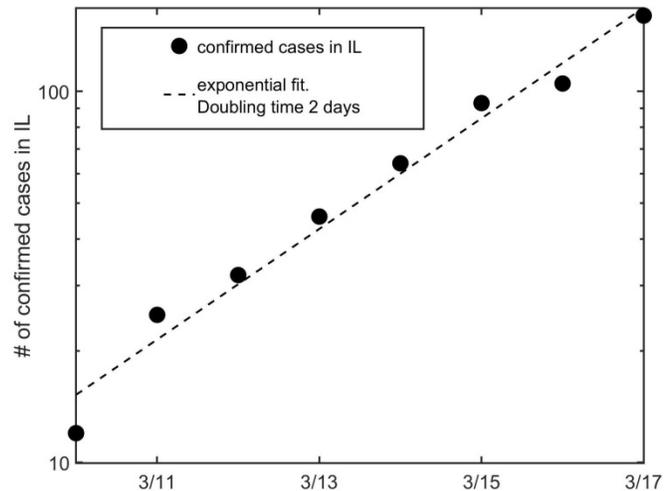

Figure 1 Number of confirmed COVID_19 cases in Illinois vs. date. The vertical axis is a logarithmic scale. The linearity of this graph shows the exponential growth with doubling time of 2 days expected early in an epidemic. The simulation model is calibrated to replicate this trend.

We ran the simulation until September 1 2020, and this was long enough to see the time course of the epidemic. The goal of our numerical experiments was to estimate the effect of strong mitigation scenarios on the peak number of severe cases (requiring hospitalization), peak number of critical cases (requiring hospitalization and special treatment) and total number of deaths. In particular, we wanted to observe the sensitivity of the outcomes to the time when strong mitigation was implemented, rather than the effects of different potential mitigations implemented at the same time [11].

Strong mitigation is defined in our calculation by reducing the transmission in such a way that the effective epidemiological parameter $R_0$ drops below unity and stays there. If $R_0 < 1$, it is a mathematical certainty that the infection will die out, and the smaller it is, the faster the die-out occurs.

|  | Peak demand for ICU beds | Peak date | Peak of hospitalizations | Total deaths by 9/1/2020 |
|---|---|---|---|---|
| Just-in-time mitigation | 128 | 22 Apr 2020 | 470 | 1151 |
| Delayed mitigation | 1585 | 29 Apr 2020 | 6012 | 7445 |

Table 1. Prediction of the number of ICU beds needed in Chicago, the date when this peak will be reached, the peak in the number of hospital beds required, and the total number of deaths by September 1, 2020. The rows correspond to two scenarios: just-in-time and delayed mitigation scenarios described in the text.

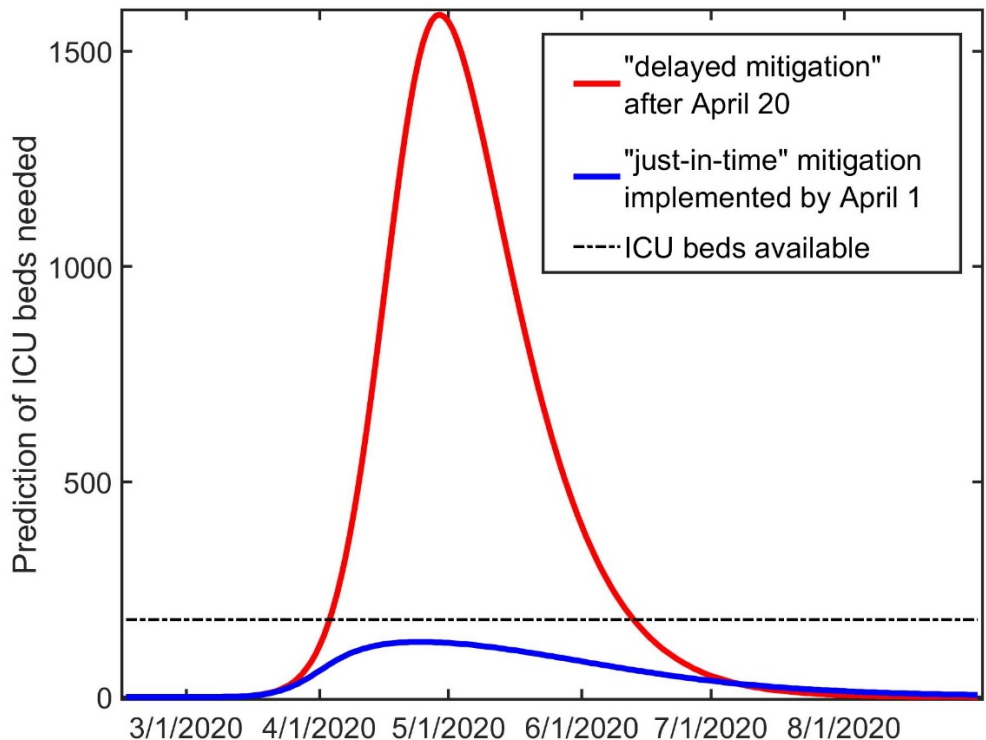

**Figure 2. Prediction of the number of ICU beds needed in Chicago vs. date.** The horizontal dashed line shows the current ICU beds (180) available in the City of Chicago on March 18 2020. It is desirable to keep the demand for ICU beds below the horizontal dashed line so that the capacity is not exceeded. The curve in red shows the results of a simulation with strong mitigation implemented after April 20th 2020. The demand exceeds the number of available ICU beds by a factor of about ten. The curve in blue shows the results if strong mitigation is implemented before April 1st 2020. In this case the demand does not exceed the number of available ICU beds.

In Figure 2 and Table 1, we show the results of simulating the effects of strong mitigation on the demand for ICU beds in Chicago. We ran two simulations with the parameter $R_0 = 0.9$, probably the least severe mitigation that one can do which is guaranteed to eventually end the COVID-19 epidemic. In one simulation (the "late mitigation" scenario), shown by the red curve, the mitigation was fully implemented by April 20th 2020. In the other, shown by the blue curve (the "just-in-time mitigation" scenario), the mitigation was fully implemented by April 1st 2020. The implementation of the mitigations was assumed to ramp up gradually over a period of time, and the strong steps already taken in Illinois are part of this mitigation and included in the calculation. Further mitigations are no doubt possible, and strong mitigation would be the final additional step.

The late mitigation scenario predicts that the number of ICU beds needed as the epidemic progresses in Chicago quickly exceeds the city's available beds (180, shown by the dashed horizontal line) and even the entire ICU bed capacity of the city (749, not shown for reasons of clarity). **The amount by which the available capacity is exceeded is by a factor of 10, and the** **amount by which the total capacity is exceeded is more than a factor of 2. In this scenario, the total number of deaths is estimated as 7445.**

The just-in-time mitigation scenario predicts that at the peak of the epidemic, the demand for ICU beds does not exceed the number of available beds, and is significantly below the city's total ICU capacity. In this scenario, the "curve is flattened", and the epidemic dies out without a catastrophic impact on the city's ability to cope, and without ICU beds having to be set aside from their regular functions, associated with surgery etc. In this scenario, the total number of deaths is estimated as 1151.

**Discussion**

Both scenarios involve strong mitigations, certainly stronger than current measures already implemented. Both mitigations end the epidemic eventually, around August. But one causes along the way a catastrophic event for the healthcare system, with many potential deaths predicted, and one does not. We note that in both Wuhan and Italy, strong mitigation measures were taken such as lockdown, but they were taken too late and still resulted in thousands of lives lost, including those of healthcare professionals. As the epidemic develops along its inevitable exponential growth trajectory, it is equally inevitable that leadership will eventually be forced to implement lockdown. Thus, if this is going to happen anyway, it should be taken as early as possible.

Our calculation shows that the just-in-time scenario accomplishes this and that the window for such a strong mitigation will soon be closed. Of course, one cannot be sure that the dates and course of the epidemic are going to follow the precise predictions we have made. There are inevitable uncertainties in making predictions for such a powerful phenomenon as an epidemic. However, whether the outcomes predicted by our scenarios occur on exactly the dates given here, and with exactly the numbers provided here, they will occur as we have predicted. The window that we have predicted is rather short, perhaps two weeks at the longest. Thus to be safe, the mitigation measures should be implemented as soon as possible, while they will be effective; at the same time, for humanitarian reasons, the societal implications of lockdown require appropriate arrangements to be made for the city's population, as has already done for the mitigation efforts to date.

**Acknowledgments**